\documentclass{aa}  
\usepackage{capt-of}
\usepackage{ragged2e}
\usepackage{graphicx}
\usepackage{threeparttable} 
\usepackage{booktabs} 
\usepackage{txfonts}
\usepackage{arydshln}
\usepackage{multirow}
\usepackage{lscape}
\usepackage{longtable}
\usepackage{booktabs}
\usepackage{xcolor} 
\usepackage[title]{appendix}
\usepackage{rotating}
\begin{document} 
\titlerunning{A deep near-infrared view of the Ophiuchus galaxy cluster}

   \authorrunning{Galdeano et al.}

   \title{A deep near-infrared view of the Ophiuchus galaxy cluster}


   \author{D. Galdeano \inst{1}\thanks{dgaldeano@unsj-cuim.edu.ar},
        G. Coldwell\inst{1},
        F. Duplancic\inst{1},
        S. Alonso\inst{1},
        L. Pereyra\inst{2},
        D. Minniti\inst{3,4},
        R. Zelada Bacigalupo\inst{5},
        C. Valotto\inst{2,6},
        L. Baravalle\inst{2},
        M.V. Alonso\inst{2,6}, 
        \and
        J.L. Nilo Castellón\inst{7,8}
          }

    \institute{Departamento de Geof\'{i}sica y Astronom\'{i}a, CONICET, Facultad de Ciencias Exactas, F\'{i}sicas y Naturales, Universidad Nacional de San Juan, Av. Ignacio de la Roza 590 (O), J5372DCS, Rivadavia, San Juan, Argentina
    \and
    Instituto de Astronom\'ia Te\'orica y Experimental (IATE, CONICET-UNC), Laprida 854, C\'ordoba, X5000BGR, Argentina
    \and
    Departamento de Ciencias F\'isicas, Facultad de Ciencias Exactas, Universidad Andres Bello, Fern\'andez Concha 700, Las Condes, Santiago, Chile
    \and
    Vatican Observatory, Vatican City State, V-00120, Italy
    \and
    North Optics Instrumentos Científicos, La Serena, Chile
    \and
    Observatorio Astron\'omico de C\'ordoba, Universidad Nacional de C\'ordoba, C\'ordoba, X5000BGR, Argentina
    \and
    Dirección de Investigación y Desarrollo, Universidad de La Serena. Av. Raúl Bitrán Nachary 1305, La Serena, Chile
    \and
    Departamento de Astronomía, Universidad de La Serena. Avenida Juan Cisternas 1200, La Serena, Chile}

   \date{Received February 15, 2022; accepted XXX, 2022}

  \abstract
   {The Ophiuchus cluster of galaxies, located at low latitudes in the direction of the Galactic bulge, has been relatively poorly studied in comparison with other rich galaxy clusters, such as Coma, Virgo, and Fornax, despite being the second brightest X-ray cluster in the sky.
}
   {Our aim is perform a study of the hidden galaxy population of the massive cluster Ophiuchus located in the Zone of Avoidance.
}
   {Deep near-infrared images and photometry from the VISTA Variables in the Vía Láctea eXtended (VVVX) survey were used to detect galaxy member candidates of the Ophiuchus cluster up to 2 Mpc from the cD galaxy 2MASX J17122774-2322108 using criteria from a past paper to select the galaxies among the foreground sources. We also perform a morphological visual classification and generate color-magnitude diagrams and density profiles. 
}
   {We identify 537 candidate galaxy members of the Ophiuchus cluster up to 2 Mpc from the cD galaxy, increasing by a factor of seven the number of reported Ophiuchus galaxies. In addition, we performed a morphological classification of these galaxy candidates finding that the fraction of ellipticals reaches more than 60 \% in the central region of the cluster. On the other hand, the fraction of spirals fraction is lower than 20 \%, remaining almost constant throughout the cluster. Moreover, we study the red sequence of galaxy member candidates and use mock catalogs to explore the density profile of the cluster, finding that the value derived from the mock catalog toward an overdense region is in agreement with the galaxy excess of the central zone of the Ophiuchus cluster.
}
   {Our investigation of the hidden population of Ophiuchus galaxies underscores the importance of this cluster as a prime target for future photometric and spectroscopic studies. Moreover the results of this work highlight the potential of the VVVX survey to study extragalactic objects in the Zone of Avoidance.} 

\maketitle

\section{Introduction}

The Ophiuchus cluster of galaxies is the second brightest galaxy cluster in the X-ray sky \citep{Million2010,werner16}  and one of the hottest clusters with a cool core \citep{fujita08,perez-torres09}. Indeed, the new all-sky images released by the eROSITA (extended ROentgen Survey with an Imaging Telescope Array) mission \citep{Predehl2020} highlight this bright X-ray cluster of galaxies located at low latitudes behind the Milky Way, particularly at the edge of the bulge. Nevertheless, it has not been as well studied as the Virgo or Coma clusters due to its low Galactic latitude (b = $9.3^o$), which hampers optical observations. Ophiuchus has mostly been studied in X-rays \citep{Johnston1981,Matsu96,ebe02}. Moreover, based on Chandra observations, \citet{Million2010} have shown a small difference, $\sim 2 \rm \ kpc$, between the position of the X-ray peak and the cD galaxy 2MASX J17122774-2322108, as well as  some strong features related to the existence of a merger in the cluster center. Further, some advances have been made at optical wavelengths, although the high dust absorption at the cluster galactic latitude provides constrains in the galaxy detection.
\citet{Durret15} (hereafter, Du15) performed an optical study of dynamical properties of 142 galaxy cluster members estimating a redshift of 0.0296, a velocity dispersion $\sigma = 954 \ \rm km \rm s^{-1}$, and a mass of $1.1 \times 10^{15}\ \rm M_{\sun}$.
Later on, \citet{Durret2018} derived the cluster galaxy luminosity function finding a substantial excess of bright galaxies compared with the Coma cluster.

Observing extragalactic sources in the Zone of Avoidance (ZOA) is a challenge. Dust and stars obstruct optical observations,  and there is a lack of information due to dust absorption, resulting in an incomplete picture of the existing galaxies. Therefore, revealing the presence of extragalactic structures behind the Milky Way, such as groups or clusters, represents an enormous observational endeavor.

However, the near-infrared public survey VISTA Variables in V\'ia L\'actea (VVV, \cite{min10,sai12} ) has led to significant advances. The pioneering work of \cite{col14} presents the VVV near-infrared galaxy counterparts of a new cluster of galaxies at redshift 0.13 observed in X-rays by SUZAKU \citep{mor13}. They detected 15 new galaxies candidates, within the central region of the cluster up to $350 \ \rm kpc$ from the X-ray peak emission, with typical  magnitudes and colors of galaxies at the cluster redshift. In addition, \cite{bar19} confirmed the existence of the first galaxy cluster, discovered by the VVV survey beyond the galactic disk, by using spectroscopic data from Flamingos-2 at GEMINI Observatory. Recently, \cite{gal21} found an unusual concentration of galaxies, in the VVV tile $b204$, detecting 624 extended sources, 607 of which are new galaxy candidates that have been cataloged for the first time. 
By exploring the spatial galaxy distribution in a small region of 15 arcmin, these authors found a noticeable overdensity, reinforcing this result with mock catalog information.

The VISTA Variables in the Vía Láctea eXtended (VVVX) survey observed a wider area around the ZOA using near-infrared images \citep{min18}. This region contains the very massive Ophiuchus galaxy cluster. In this work we aim to study, for the first time, the galaxy population of the Ophiuchus cluster by using near-IR images.  We obtain data of several new galaxy member candidates in the Ophiuchus cluster, morphologically classify the galaxies, and analyze the morphology-density relation and the density profiles around the cluster.

The layout of the paper is as follows: The observational data, galaxy candidate detections and morphological classification are described in Sect. 2. In Sect. 3 the morphology-density relation and the density profiles around the cluster are analyzed. The conclusions are presented
in Sect. 4. Throughout the paper, we use the following cosmological parameters: 
$H_0 = 70\ \rm km\ \rm s^{-1}\ \rm h^{-1}$, $\Omega_m = 0.3$, and $\Omega_\lambda = 0.7$.

\section{Observational data}
\subsection{VVVX survey}
\label{vvvxsurvey}

The VVVX survey is an extension of the VVV Galactic near-infrared European Southern Observatory (ESO) survey, covering more areas of the Milky Way bulge and disk \citep{min18}. In the Galactic bulge, the VVVX provides a spatial coverage of about 1700 deg$^{2}$, using the J ($1.25\ \mu$m), H ($1.64\ \mu$m ), and Ks ($2.14\  \mu$m) near-infrared passbands with the VISTA InfraRed CAMera at the $4.1$ m ESO Visual and Infrared Survey Telescope for Astronomy (VISTA ; \citet{Emer04, Emer06, Emer10}), located at Cerro Paranal in  Chile.

The VVVX observations were reduced with the VISTA Data Flow System \citep{Irwin04, Emer04} at the Cambridge Astronomical Survey Unit (CASU\footnote{http://casu.ast.cam.ac.uk/surveys-projects/vista}; \citep{lewis2010}).  The catalogs and processed images are available from the ESO Science Archive\footnote{http://archive.eso.org/} and from the VISTA Science Archive\footnote{http://horus.roe.ac.uk/vsa/} (VSA; \citep{Cross12}).

The Ophiuchus cluster, at redshift z $\approx$ 0.03 (Du15), is located in the ZOA, behind the Galactic bulge and near the edge of the VVVX survey area.
The cD galaxy 2MASX J17122774-2322108 is taken as the cluster center position ($\rm  RA=258^{\circ}.1155$, $\rm Dec =-23^{\circ}.3698$, $J2000.0$), very close to the X-ray  peak emission \citep{Million2010}. It is located in tile b506, 10 arcmin from the survey border (corresponding to $\sim 360$ kpc considering the cluster redshift). Furthermore, the galaxy cluster region also covers part of the b505 tile. Therefore, for the present work we downloaded tile-stack images of both the b505 and b506 tiles. To this end, we used the VSA \texttt{GetImage} tool and selected J, H, and Ks images observed the same night in order to have a similar sky quality for the observations in the different bands.

\subsection{Galaxy detection and classification}

The multiwavelength catalog used in this work was built by using SExtractor \citep{ber96} on the VVVX images from tiles b505 and b506. The output parameters selected from SExtractor catalogs were the equatorial coordinates and the J, H, and Ks magnitudes in three-pixel-radius aperture. They were used to calculate the galaxy colors and total magnitudes given by $MAG\_AUTO$, which is based on Kron's algorithm \citep{kro80}. The magnitudes were extinction-corrected with extinction maps adopted from \cite{Schlafly2011}. Also, we selected the parameters $CLASS\_STAR$ (CS) and $half-light$ radius ($r_{1/2}$), used for $star-galaxy$ separation.

In order to obtain the galaxy candidates, we used the $star-galaxy$ separation parameters considering $r_{1/2} > 0.7 \rm \ arcsec$, $CS<0.5$, and magnitude cut 10$<$Ks$<$16.5 mag, following the criteria from \cite{gal21}. The lower limit was selected in order to prevent saturated objects, and the upper limit represents a good approach to the reliability limit for visual inspection \citep{gal21}. Under these constrains, we identified 4168 extended sources. Additionally, we performed a visual inspection of these sources in order to discard false detections and selected objects with morphological features and surface brightnesses resembling galaxies. Thereafter we built a catalog of 537 candidate galaxy members of the Ophiuchus cluster up to 2 Mpc from the cD galaxy.

In addition, the morphological features of the galaxy candidates observed in the VVVX images allowed us to classify them, as ellipticals (with bulge-type appearance) or spirals (with a noticeable disk), or as uncertain when the morphology is not clear enough. It is important to notice that the uncertain galaxies are, in general, weak or low surface brightness sources, which prevents a confident classification. 

\begin{figure*}[htb]
\centering
    \includegraphics[width=1.6\columnwidth]{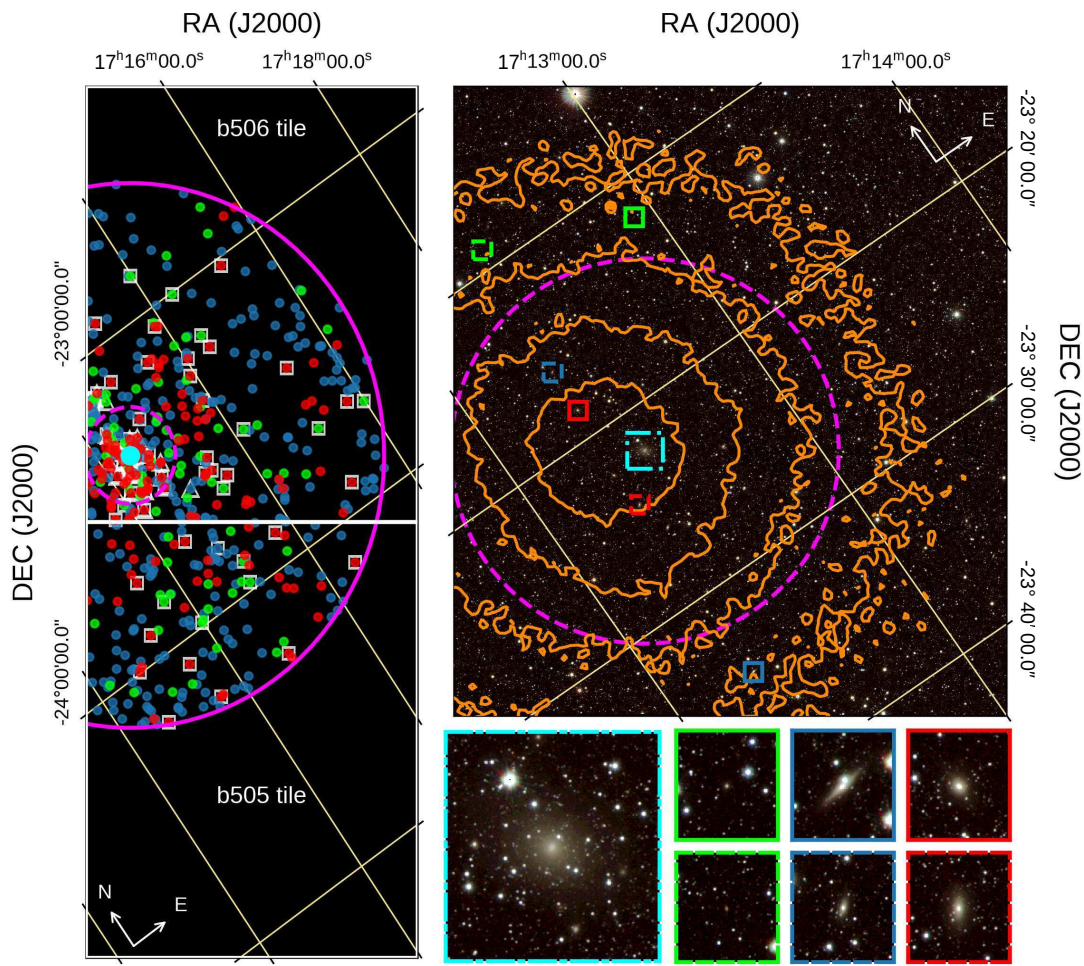}
    \caption{Left: Schematic view of the b505 and b506 tiles, with the inner and outer regions indicated with dashed and solid magenta circles, respectively. Both the new galaxies cataloged in this work and those from the literature are included. The cyan dot corresponds to the cD galaxy. The red, blue, and green filled circles represent the elliptical, spiral, and uncertain galaxies, respectively, that have been detected and classified in this paper. The white squares correspond to \cite{Durret15} galaxies and the white triangles to 2MASX galaxies. 
    Right: False-color J (blue), H (green), and Ks (red) images of the inner region of the Ophiuchus galaxy cluster located in the VVVX b506 tile. 
    The solid contours (orange) delineate the X-ray map drawn from ASCA data from the central cluster region. The colored boxes (zoomed-in views in the bottom-right panels) show examples of galaxy candidates that correspond to the different morphological classifications. 
    The cyan box corresponds to the cD galaxy and the red, blue, and green boxes to the elliptical, spiral, and uncertain galaxies, respectively.
    The smaller boxes have lengths of 30 arcseconds, while the side of the larger box is approximately 1 arcminute.
    }
    \label{cluster}
\end{figure*}

Figure \ref{cluster} (right) shows the false-color red-green-blue (RGB) image of the inner region of Ophiuchus galaxy cluster within an area of $28 \rm \times 32 \rm$ [arcmin].
The contours based on an X-ray map drawn from Advanced Satellite for Cosmology and Astrophysics (ASCA) data \citep{wat01} delineate the central region. 
Zoomed-in views of the cD galaxy and six new detected galaxies, two for each morphological classification, are shown in the lower-right panels Fig. \ref{cluster}. In the right panel of this figure we can also observe a difference between the cD galaxy position with respect to the center of the X-ray contours. This offset was detected by \cite{perez-torres09} and estimated to be in $\sim 2 kpc$, suggesting that the central region of the Ophiuchus cluster is in the process of merging.

The full catalog of galaxy candidates is available at the Centre de Données astronomiques de Strasbourg (CDS; see Appendix A).

\begin{table}
\centering
\caption{Number of extragalactic candidates with 10$<$Ks$<$16.5 mag for the inner, outer, and whole region.}
 \begin{threeparttable}[t]
\vskip 0.1cm
\begin{tabular}{| l | ccc | c | c | c |}
\hline
\rule{0pt}{3ex} \rule[-1.5ex]{0pt}{0pt} & \multicolumn{4}{|c|} {This Work} &  Du15\tnote{4} & 2MASX\tnote{4} \\
\hline \hline
\rule{0pt}{3ex} \rule[-1.5ex]{0pt}{0pt}
Region & E & S & U & Total & Total & Total\\
\hline 
\rule{0pt}{3ex} \rule[-1.5ex]{0pt}{0pt}
inner\tnote{1}
& 51 & 16 & 38 & 105 & 27 & 20\\
\hline
\rule{0pt}{3ex} \rule[-1.5ex]{0pt}{0pt}
outer\tnote{2}
& 103 & 64 & 265 & 432 & 47 & 54\\
\hline  
\rule{0pt}{3ex} \rule[-1.5ex]{0pt}{0pt}
whole\tnote{3}
& 154 & 80 & 303 & 537 & 74 & 74\\
\hline 
\end{tabular}
\begin{tablenotes}
     \item[1] Area with $\rm r_p < 360\ \rm kpc$ ($\sim 0.09 \rm \ deg^2$). 
     \item[2] Area  between $360\ \rm kpc < \rm r_p <2\ \rm Mpc$. 
     \item[3] Area with $\rm r_p <2\ \rm Mpc$ ($\sim 1.75 \rm \ deg^2$).
     \item[4] Number of galaxies detected by Du15 and 2MASX listed in each region.
   \end{tablenotes}
    \end{threeparttable}
\label{tt}
\end{table}

\section{Results}
The position of the cluster center on tile b506 is located 10 arcmin from the VVVX survey boundary, meaning the projected distance is 360 kpc at the cluster redshift. This allows us to perform a symmetrically complete study of the cluster center area (inner region).
For projected distances $\rm r_p > 360\ \rm kpc$ (outer region),
the analysis is extended up to $2\ \rm Mpc$ for comparison with existing catalogs. A schematic view of the defined inner and outer regions, including the galaxies detected in this work, Du15, and the 2MASS Extended Source Catalog (2MASX), is shown in Fig. \ref{cluster} (left panel).

The number of sources detected in this work within the inner, outer,  and $\rm r_p < 2\ \rm Mpc$ regions are listed in Table 1. The proportion of elliptical, spiral and uncertain galaxies is also presented for the different galaxy cluster regions. We note that these spatial scales are similar to those of the Coma cluster, a rich cluster of galaxies located in the northern hemisphere \citep{Hammer2010,Poloji2022}.

\subsection{The galaxy cluster population}

The works of Du15 and \cite{Jarrett2000} based on optical and near-infrared data, respectively, can be compared to each other fairly well, allowing the improvement obtained from using VVVX near-infrared data to be quantified. The total numbers of cataloged galaxies from the mentioned papers, including those in the defined inner, outer, and whole regions, are also listed in Table 1. From this table we observe that the numbers of galaxy candidates detected in this work are four and eight times higher than those from Du15 and 2MASX for the inner and outer region, respectively. Moreover, if we consider the whole region, we find $\sim$ 7 times more galaxies within  $\rm r_p < 2 \rm Mpc$ compared to previous works. This result indicates that VVVX provides a meaningful enhancement of galaxy detection allowing a more complete picture of the content of galaxies of the Ophiuchus cluster.

\begin{figure}[htb]
    \centering
    \includegraphics[width=0.99\columnwidth]{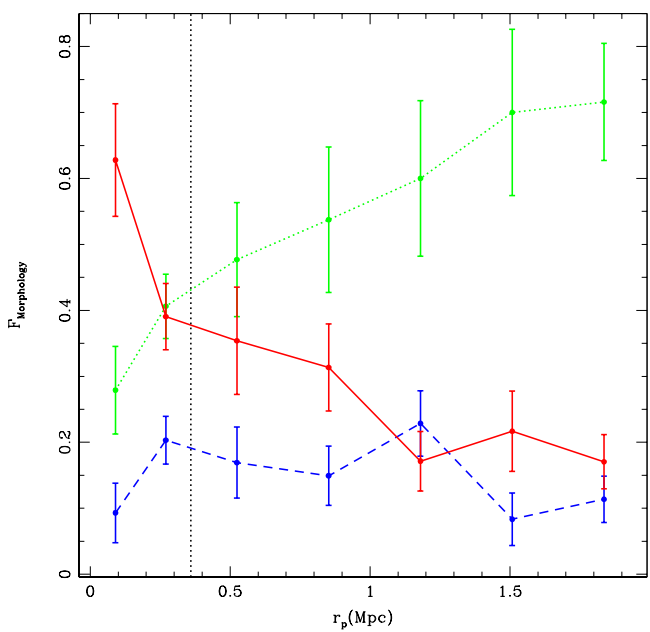}
    \caption{Fraction of morphological classified galaxies as a function of the projected distance to the cD galaxy. Solid red, dashed blue, and dotted green lines correspond to elliptical, spiral and uncertain galaxies, respectively. The vertical dotted black line indicates the boundary between the defined inner and outer regions.}
    \label{fracMD}
\end{figure}

We also analyzed the distribution of galaxies according to their morphological classification. In Fig. \ref{fracMD} we show the fraction of elliptical, spiral, and uncertain galaxies as a function of the projected distance to the cluster center, given by the cD galaxy position. In order to avoid survey edge effects, for projected distances higher than 360 kpc (outer region) we considered a galactic latitude cut of $b<9.7^o$ (i.e., the latitude of the cD galaxy). This goal of this restriction was to have symmetry in the normalized area (see the left panel of Fig. \ref{cluster}). All the uncertainties were derived via a bootstrap resampling technique \citep{Barrow}.  

In Fig. \ref{fracMD} we observe that the proportion of elliptical galaxies is higher than 60 \% in the very central region of the cluster. On the other hand, the fraction of spiral galaxies is lower than 20 \%, remaining almost constant up to 2 Mpc. The fraction of uncertain galaxies increases significantly for the outer region as expected for these low luminosity galaxies. The observed trends are in good agreement with those expected for the morphology-density relation \citep{dre80}.

\subsection{Red sequence} 

Further, considering that galaxy clusters contain a well-defined sequence in the color-magnitude diagram \citep{gla00}, mainly composed of elliptical and lenticular galaxies, we studied the red sequence (RS) for the detected sources. We focused on the inner region to reduce the relative contamination due to field galaxies \citep{Treu03,stott2009}.

\begin{figure*}[htb]
    \centering
      \includegraphics[width=0.99\textwidth]{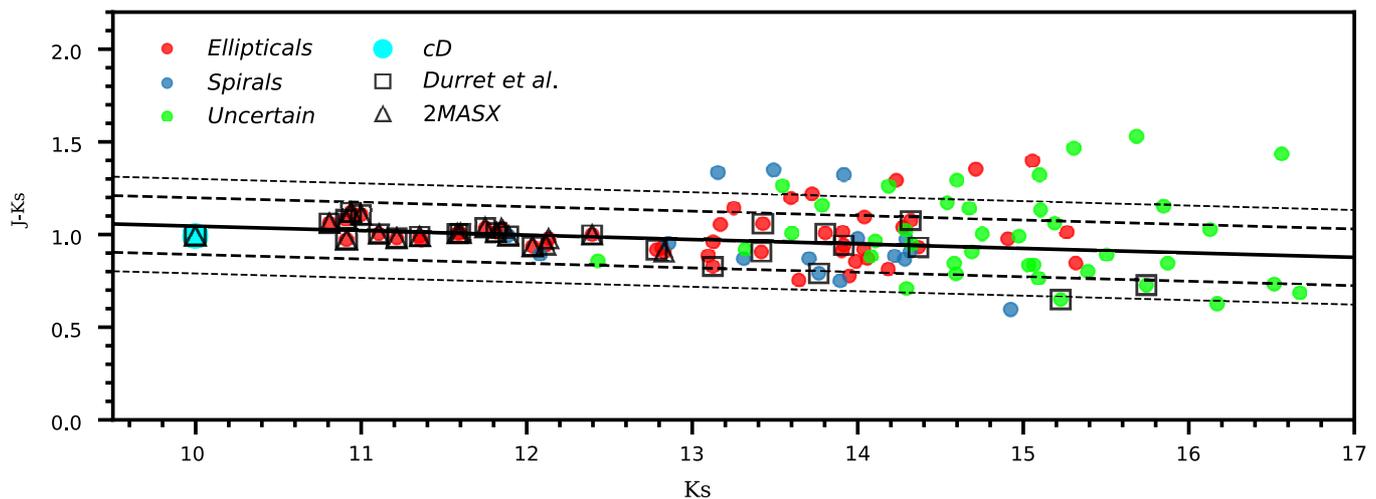}
    \caption{Near-infrared color–magnitude diagram of galaxies, J-Ks vs. Ks total magnitude. The RS fit, using the ten brightest galaxies, is shown by the solid black line. The dashed and dotted lines indicate $\pm 3\sigma$ and $\pm 5\sigma$ around the fit, respectively. The new cataloged galaxies are presented, color-coded by morphological classification. Galaxies from Du15 and 2MASX, with redshifts similar to that of the Ophiuchus cluster, are shown with open squares and open triangles, respectively.
}
    \label{RS}
\end{figure*}

In Fig. \ref{RS} we show the color-magnitude diagram for the infrared galaxies detected in this work, taking the adopted morphological classification schema into account. We also show the 2MASX galaxies and the galaxies detected by Du15. We can observe that the RS mainly comprises bright elliptical galaxies. In order to study the characteristics of the RS, we performed a linear fit using the ten brightest galaxies. Within $\pm 3 \sigma$ around the RS fit, we have 72 galaxies, and thus almost 70 \% of the galaxies belong to the inner region. If we take $\pm 5 \sigma$ around the RS fit we have 88 galaxies (84 \%). The obtained value for the slope, $k_{JKs}$ $= -0.024 \pm 0.051$, is consistent with that found by \cite{stott2009} for galaxy clusters at the redshift of Ophiuchus.

\subsection{Comparison with observational and simulated data}

The final catalog comprises 537 candidate cluster members, identified by adopting the \citet{gal21} selection criteria and using visual confirmation and morphological classification. Hereafter, we are able to compare our findings with those predicted by numerical simulations.

Following the technique developed in \citet{gal21}, we built a simulated galaxy catalog from the semi-analytical model of galaxy formation L-Galaxies \citep{Henriques2012}. This synthetic galaxy catalog provides several galaxy properties, such as  the Sloan Digital Sky Survey (SDSS) absolute magnitudes, sky position, star formation histories, stellar masses, and galaxy type. Our aim is to estimated the expected integrated mean number of galaxies  up to the cluster redshift. Therefore, in order to obtain VVV magnitudes from the SDSS magnitudes, we first used the transformations proposed by \citet{bilir2008} and then those presented in \citet{GonzalezFernandez2018}. To calculate the apparent magnitudes in the observer frame, we estimate the observational redshift using the peculiar velocity of each galaxy in the direction of the line of sight. Then, we applied inverse K-corrections following the \citet{Chilingarian2010} model. Since we are trying to simulate a region very close to the bulge of the Milky Way, we used the \citet{Schlafly2011} dust extinction model to correct these magnitudes.

Finally, we used the mock catalog to calculate the galaxy counts by directing the line of sight toward both a random and an overdense region. It is worth mentioning that, for the mock catalog galaxies, we applied cuts in apparent magnitude $10< \mathrm{Ks} < 17$ and color $\mathrm{J-Ks} > 0.97$, $\mathrm{J-H} > 0$, and $\mathrm{H-Ks}>0$ in order to have photometric properties similar to those of the observational galaxy sample.

\begin{figure}[htb]
    \centering
    \includegraphics[width=1.0\columnwidth]{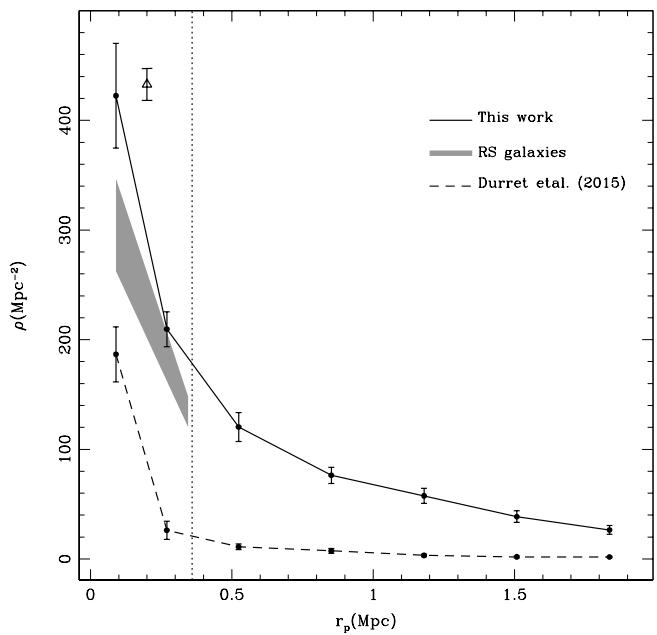}
    \caption{Density profiles as a function of the projected distance to the cD galaxy for the whole galaxy sample from this work, the galaxies from the inner region (within $3\sigma$ of the RS linear fit) from this work, and the galaxy sample taken from Du15. The vertical dotted black line indicates the boundary between the defined inner and outer regions.
    The open triangle represents the predicted value from the mock catalog.}
    \label{ro}
\end{figure}

Thus, according to the mock results the mean background expected value, within 1 $\rm Mpc^{2}$, is  23.2 $\pm$ 3.1 galaxies. Regarding overdense regions, the mock catalog predicts we will find 432.7 $\pm$ 9.6 mock galaxies per $\rm Mpc^2$; this is in agreement with the excess of galaxies achieved in the central region of the galaxy cluster, which can be observed in the density profile from Fig. \ref{ro}.

Density profiles from Fig. \ref{ro} have been estimated considering projected distances larger than 360 kpc (outer region). To have symmetry in the normalized area, the analysis was restricted to objects with galactic latitudes lower than that of the cD galaxy ($b<9.7^o$). In this figure we show density profiles as a function of the projected distance to the cD galaxy for the whole galaxy sample and the galaxies from the inner region within $3\sigma$ of the RS linear fit from this work. For comparison, we also plot the density profile from the galaxy sample taken from Du15. From this figure we can appreciate that there is an excess of the density profiles of the galaxy member candidates obtained in this work in the inner region with respect to the galaxy sample taken from Du15 (two times more). In addition, the profile calculated with the galaxies of the RS also shows a clear excess in comparison to that obtained by Du15.

\section{Summary and conclusions}

We have investigated the hidden galaxy population of the massive Ophiuchus cluster. We used deep near-infrared photometry provided by the VVVX survey and analyzed the b505 and b506 tiles, over which the cluster extends. These tiles are located behind the Galactic bulge, near the edge of the VVVX.

We used the software SExtractor to generate a multi-band catalog with stellar and extended sources using the star-galaxy classification parameters CS and $r_{1/2}$. Then, following \cite{gal21}, we considered objects with $CS < 0.5$ and $r_{1/2} > 0.7$ arcsec as likely extended. We also adopted the magnitude cut $10 < Ks < 16.5$ mag and performed a detailed visual inspection of false-color RGB images constructed from filters J, H, and Ks.

Following this procedure, we obtained a total of 537 candidate cluster members of the Ophiuchus galaxy cluster up to 2 Mpc from the cD galaxy.
This result represents $\approx$ 4 and $\approx$ 8 times more cluster galaxy candidates in the inner ($\rm r_p < 360\ \rm kpc$) and outer ($360\ \rm kpc < \rm r_p < 2\ \rm Mpc$) regions, respectively, than the fraction obtained in previous works (Du15 and 2MASX).

We also classified the galaxy candidates following the morphological features as elliptical and spiral, or uncertain when the morphology is not clear enough. We found that the proportion of ellipticals reaches more than the 60 \% in the very central region of the cluster. On the other hand, the fraction of spirals is lower than 20 \%, remaining almost constant up to 2 Mpc. The fraction of uncertain galaxies increases significantly for the outer region, as expected for these low luminosity galaxies. This result is in good agreement with that expected from the morphology-density relation \citep{dre80}, indicating the suitable morphological classification obtained from the VVVX images in this work.

In addition, we studied the sample of galaxy member candidates of the Ophiuchus cluster using the RS.
We performed a linear fit with the ten brightest galaxies, finding that within $\pm 3 \sigma$ and $\pm 5 \sigma$ around the RS fit we get a fraction of 70 \% and 84 \% of galaxies respectively. Also the obtained value for the slope is consistent with that found by \cite{stott2009} for galaxy clusters at the redshift of Ophiuchus.

We constructed a simulated galaxy catalog in order to estimate the total number of galaxies expected to be found up to the Ophiuchus redshift. The galaxy counts obtained by directing the line of sight toward a mock overdense region are in agreement with the excess of galaxies detected in the central area of the Ophiuchus cluster. We calculated the density profile using the RS galaxy member candidates, finding twice as many sources within the inner cluster region than in the galaxy sample of Du15. The high number of elliptical galaxies agrees with the scenario proposed by different authors regarding the characteristics of the central population of the Ophiuchus cluster. By analyzing the X-ray emission of the cluster, \cite{Million2010} found a comet-like morphology, which indicates that the cluster is merging. This morphology may be produced by ram-pressure stripping, a rapid motion through the intracluster medium. This is a common process that describes the removal of a galaxy gaseous components propitiating the movement of galaxies from the blue regions in the color magnitude space to the area where the RS is located (\cite{Hughes2009}, and references therein). High-resolution spectroscopy or even integral field unit spectroscopy of these central galaxies will be crucial in confirming the impact of the environment on the internal physical properties of the galaxies, helping shed light on the formation of red-cluster sequence galaxies. 

Our investigation of the Ophiuchus cluster using VVVX data to study the hidden population of galaxies in the ZOA demonstrates the potential of this survey and underscores the importance of this galaxy cluster.
Being the nearest massive cool-core cluster in the southern hemisphere, it is a prime target for future spectroscopic and photometric studies.
Deep high-resolution imaging that could become available with the Vera Rubin Telescope (LSST Science Collaboration 2009)would allow a more accurate classification of galaxies as well as the identification of fainter members, nucleated galaxies, galactic streams, and globular cluster systems. Likewise, multi-object spectroscopy could be enabled with the 4-meter Multi-Object Spectroscopic Telescope at VISTA \citep{deJong2019}, enabling a detailed chemical and dynamical characterization of the Ophiuchus cluster galaxies.

\begin{acknowledgements}
D.G. PhD studentship is supported by the UNSJ. This work was partially supported by the Consejo Nacional de Investigaciones Cient\'{\i}ficas y T\'ecnicas and the Secretar\'{\i}a de Ciencia y T\'ecnica de la Universidad Nacional de San Juan. 
The authors gratefully acknowledge data from the ESO Public Survey program ID 179.B-2002 taken with the VISTA telescope, and products from the Cambridge Astronomical Survey Unit (CASU).D.M. gratefully acknowledges support by the ANID BASAL projects ACE210002 and FB210003, and Fondecyt Project No. 1220724. MVA and CV acknowledge the support of the Consejo de Investigaciones
Científicas y Técnicas (CONICET) and Secretaría de Ciencia y Técnica de la Universidad Nacional de Córdoba (SECyT). J. L. N. C. is grateful for the financial support received from the Southern Office of Aerospace Research and Development of the Air Force Office of the Scientific Research International Office of the United States (SOARD/AFOSR) through grants FA9550-18-1-0018 and FA9550-22-1-0037. 

   \end{acknowledgements}

%
   \bibliographystyle{aa} 
  \bibliography{references} 
%


\begin{appendix}
\section{Ophiuchus galaxy catalog}
\label{A1}
The final catalog (Table A.1) lists the photometric and morphological parameters of the galaxies identified in this work. It contains the galaxy name in column (1), the J2000 equatorial coordinates in columns (2) and (3), the photometric data (three-pixel aperture and total magnitudes) in columns (4) to (9), the CS and r$_{1/2}$ parameters in columns (10) and (11), visual morphological classification in column (12), and the redshift information obtained from the existing catalogs with the corresponding references when available in columns (13) to (15). The full catalog is available at the CDS.

\begin{landscape}

\begin{flushleft}
\captionof{table}{Table A.1. A sample of 40 randomly selected galaxies in our catalog. The full table is available at the CDS. See appendix A for a detailed description.} 
\end{flushleft}

\scriptsize
 \begin{tabular}{lllllllllllllll}
\hline
\hline
VVVX & RA (J2000) & Dec (J2000) & Ks(3) & H(3) & J(3) & Ks & H & J & CS & $r_{1/2}$ & MT & z  &  Reference & Bibcode \\ 
\hline
name & hh:mm:ss & dd:mm:ss & mag & mag & mag & mag & mag & mag &  & arcsec &  &   &  &  \\ 
\hline
  VVVXJ171004.21-240242.5 & 17:10:04.00 & -24:02:43.0 & 15.95$\pm$0.05 & 16.25$\pm$0.04 & 16.77$\pm$0.05 & 15.04$\pm$0.05 & 15.27$\pm$0.03 & 15.80$\pm$0.04 & 0.01 & 1.13 & U &  &  & \\
  VVVXJ171007.81-240116.9 & 17:10:08.00 & -24:01:17.0 & 16.33$\pm$0.06 & 16.65$\pm$0.05 & 17.35$\pm$0.07 & 14.64$\pm$0.05 & 15.07$\pm$0.03 & 15.82$\pm$0.04 & 0.00 & 1.93 & U &  &  & \\
  VVVXJ171029.70-240219.9 & 17:10:30.00 & -24:02:20.0 & 15.32$\pm$0.03 & 15.85$\pm$0.03 & 16.58$\pm$0.05 & 14.32$\pm$0.03 & 14.68$\pm$0.02 & 15.27$\pm$0.03 & 0.08 & 1.31 & S &  &  & \\
  VVVXJ171037.68-240430.0 & 17:10:38.00 & -24:04:30.0 & 16.37$\pm$0.06 & 16.45$\pm$0.04 & 17.10$\pm$0.06 & 15.65$\pm$0.06 & 15.52$\pm$0.03 & 16.12$\pm$0.04 & 0.20 & 0.99 & U &  &  & \\
  VVVXJ171038.98-235819.3 & 17:10:39.00 & -23:58:19.0 & 16.09$\pm$0.05 & 17.09$\pm$0.06 & 17.62$\pm$0.08 & 15.03$\pm$0.05 & 15.58$\pm$0.04 & 16.26$\pm$0.05 & 0.20 & 1.45 & U &  &  & \\
  VVVXJ171048.51-235900.5 & 17:10:49.00 & -23:59:01.0 & 13.81$\pm$0.02 & 14.11$\pm$0.01 & 14.91$\pm$0.02 & 12.22$\pm$0.01 & 12.42$\pm$0.01 & 13.26$\pm$0.01 & 0.03 & 2.26 & E &  &  & \\
  VVVXJ171059.88-234649.9 & 17:11:00.00 & -23:46:50.0 & 15.20$\pm$0.03 & 15.87$\pm$0.03 & 16.79$\pm$0.05 & 14.33$\pm$0.03 & 14.76$\pm$0.02 & 15.49$\pm$0.03 & 0.04 & 1.16 & U &  &  & \\
  VVVXJ171104.55-240906.6 & 17:11:05.00 & -24:09:07.0 & 16.45$\pm$0.07 & 17.00$\pm$0.06 & 17.67$\pm$0.08 & 15.73$\pm$0.08 & 16.25$\pm$0.05 & 17.03$\pm$0.07 & 0.14 & 0.98 & U &  &  & \\
  VVVXJ171120.54-235455.9 & 17:11:21.00 & -23:54:56.0 & 15.64$\pm$0.04 & 16.08$\pm$0.04 & 16.84$\pm$0.05 & 13.79$\pm$0.03 & 14.08$\pm$0.02 & 14.74$\pm$0.02 & 0.00 & 2.44 & S &  &  & \\
  VVVXJ171125.68-241112.2 & 17:11:26.00 & -24:11:12.0 & 16.24$\pm$0.06 & 16.60$\pm$0.05 & 17.28$\pm$0.07 & 15.35$\pm$0.05 & 15.55$\pm$0.03 & 16.11$\pm$0.04 & 0.00 & 1.11 & U &  &  & \\
  VVVXJ171131.85-240750.8 & 17:11:32.00 & -24:07:51.0 & 16.37$\pm$0.06 & 16.84$\pm$0.05 & 17.56$\pm$0.08 & 15.64$\pm$0.07 & 15.97$\pm$0.05 & 16.64$\pm$0.06 & 0.25 & 1.00 & U &  &  & \\
  VVVXJ171147.17-232807.1 & 17:11:47.00 & -23:28:07.0 & 15.91$\pm$0.05 & 16.39$\pm$0.04 & 17.08$\pm$0.06 & 15.33$\pm$0.04 & 15.78$\pm$0.04 & 16.43$\pm$0.05 & 0.35 & 0.85 & E &  &  & \\
  VVVXJ171155.50-231608.6 & 17:11:56.00 & -23:16:09.0 & 15.01$\pm$0.03 & 15.54$\pm$0.03 & 16.21$\pm$0.04 & 13.59$\pm$0.02 & 14.61$\pm$0.02 & 15.08$\pm$0.03 & 0.02 & 1.96 & E &  &  & \\
  VVVXJ171159.27-232310.9 & 17:11:59.00 & -23:23:11.0 & 16.66$\pm$0.08 & 16.83$\pm$0.05 & 17.29$\pm$0.07 & 16.17$\pm$0.09 & 16.24$\pm$0.05 & 16.88$\pm$0.06 & 0.40 & 0.82 & U &  &  & \\
  VVVXJ171200.87-235845.6 & 17:12:01.00 & -23:58:46.0 & 15.69$\pm$0.04 & 16.03$\pm$0.03 & 16.68$\pm$0.05 & 14.02$\pm$0.03 & 14.39$\pm$0.02 & 14.78$\pm$0.02 & 0.00 & 2.15 & S &  &  & \\
  VVVXJ171201.64-241557.4 & 17:12:02.00 & -24:15:57.0 & 15.90$\pm$0.05 & 16.34$\pm$0.04 & 17.03$\pm$0.06 & 15.03$\pm$0.05 & 15.48$\pm$0.03 & 16.12$\pm$0.04 & 0.04 & 1.15 & U &  &  & \\
  VVVXJ171209.88-231424.2 & 17:12:10.00 & -23:14:24.0 & 14.91$\pm$0.03 & 15.22$\pm$0.02 & 15.92$\pm$0.04 & 13.60$\pm$0.02 & 13.53$\pm$0.01 & 14.15$\pm$0.02 & 0.18 & 1.97 & U &  &  & \\
  VVVXJ171216.18-234031.6 & 17:12:16.00 & -23:40:32.0 & 15.27$\pm$0.03 & 15.50$\pm$0.03 & 16.07$\pm$0.04 & 13.92$\pm$0.03 & 14.13$\pm$0.02 & 14.45$\pm$0.02 & 0.01 & 1.72 & S &  &  & \\
  VVVXJ171218.01-232937.6 & 17:12:18.00 & -23:29:38.0 & 16.97$\pm$0.10 & 17.32$\pm$0.07 & 17.65$\pm$0.08 & 16.67$\pm$0.12 & 17.02$\pm$0.07 & 17.12$\pm$0.07 & 0.19 & 0.71 & U &  &  & \\
  VVVXJ171218.98-232219.1 & 17:12:19.00 & -23:22:19.0 & 12.91$\pm$0.01 & 13.21$\pm$0.01 & 13.92$\pm$0.01 & 11.11$\pm$0.00 & 11.25$\pm$0.00 & 11.88$\pm$0.01 & 0.03 & 2.78 & E & 0.028 & 2MASX J17121895-2322192 & 2020yCat.1350....0G\\
  VVVXJ171219.26-233439.7 & 17:12:19.00 & -23:34:40.0 & 12.60$\pm$0.01 & 12.92$\pm$0.01 & 13.62$\pm$0.01 & 10.79$\pm$0.00 & 11.03$\pm$0.00 & 11.79$\pm$0.01 & 0.03 & 2.97 & S &  &  & \\
  VVVXJ171220.92-231610.8 & 17:12:21.00 & -23:16:11.0 & 13.25$\pm$0.01 & 13.53$\pm$0.01 & 14.25$\pm$0.02 & 11.89$\pm$0.01 & 11.97$\pm$0.01 & 12.61$\pm$0.01 & 0.03 & 2.16 & S & 0.024 & 2MASX J17122092-2316108 & 2020yCat.1350....0G\\
  VVVXJ171226.73-232703.4 & 17:12:27.00 & -23:27:03.0 & 14.41$\pm$0.02 & 14.98$\pm$0.02 & 15.74$\pm$0.03 & 13.15$\pm$0.02 & 13.59$\pm$0.01 & 14.28$\pm$0.02 & 0.03 & 1.56 & S &  &  & \\
  VVVXJ171229.48-232123.4 & 17:12:29.00 & -23:21:23.0 & 14.86$\pm$0.03 & 15.36$\pm$0.02 & 16.08$\pm$0.04 & 13.72$\pm$0.02 & 14.10$\pm$0.02 & 14.40$\pm$0.02 & 0.03 & 1.56 & E &  &  & \\
  VVVXJ171231.05-240551.9 & 17:12:31.00 & -24:05:52.0 & 16.24$\pm$0.06 & 16.78$\pm$0.05 & 17.47$\pm$0.07 & 14.85$\pm$0.05 & 15.79$\pm$0.04 & 16.19$\pm$0.05 & 0.01 & 1.94 & U &  &  & \\
  VVVXJ171241.45-240115.6 & 17:12:41.00 & -24:01:16.0 & 15.55$\pm$0.04 & 16.00$\pm$0.03 & 16.72$\pm$0.05 & 14.90$\pm$0.04 & 15.23$\pm$0.03 & 15.78$\pm$0.04 & 0.08 & 0.93 & U &  &  & \\
  VVVXJ171242.41-241325.5 & 17:12:42.00 & -24:13:26.0 & 14.89$\pm$0.03 & 15.72$\pm$0.03 & 16.64$\pm$0.05 & 13.76$\pm$0.02 & 14.09$\pm$0.02 & 14.50$\pm$0.02 & 0.03 & 1.48 & S &  &  & \\
  VVVXJ171247.26-234042.9 & 17:12:47.00 & -23:40:43.0 & 14.43$\pm$0.02 & 14.69$\pm$0.02 & 15.38$\pm$0.03 & 13.28$\pm$0.02 & 13.40$\pm$0.01 & 14.00$\pm$0.02 & 0.03 & 1.59 & E &  &  & \\
  VVVXJ171248.05-232708.5 & 17:12:48.00 & -23:27:09.0 & 14.61$\pm$0.02 & 14.82$\pm$0.02 & 15.51$\pm$0.03 & 12.83$\pm$0.01 & 12.93$\pm$0.01 & 13.55$\pm$0.01 & 0.02 & 2.34 & E &  &  & \\
  VVVXJ171306.80-225535.1 & 17:13:07.00 & -22:55:35.0 & 13.49$\pm$0.01 & 13.79$\pm$0.01 & 14.50$\pm$0.02 & 11.60$\pm$0.01 & 11.76$\pm$0.01 & 12.41$\pm$0.01 & 0.03 & 3.50 & E & 0.024 & 2MASX J17130678-2255350 & 2020yCat.1350....0G\\
  VVVXJ171308.04-234720.6 & 17:13:08.00 & -23:47:21.0 & 15.24$\pm$0.03 & 15.68$\pm$0.03 & 16.37$\pm$0.04 & 14.35$\pm$0.03 & 14.65$\pm$0.02 & 15.24$\pm$0.03 & 0.02 & 1.15 & S &  &  & \\
  VVVXJ171310.47-234110.9 & 17:13:10.00 & -23:41:11.0 & 14.20$\pm$0.02 & 14.53$\pm$0.02 & 15.17$\pm$0.02 & 13.06$\pm$0.02 & 13.13$\pm$0.01 & 13.68$\pm$0.01 & 0.03 & 1.49 & E &  &  & \\
  VVVXJ171313.57-232511.6 & 17:13:14.00 & -23:25:12.0 & 15.92$\pm$0.05 & 16.10$\pm$0.04 & 16.74$\pm$0.05 & 14.37$\pm$0.04 & 14.41$\pm$0.02 & 14.63$\pm$0.02 & 0.00 & 1.89 & S &  &  & \\
  VVVXJ171316.72-231230.1 & 17:13:17.00 & -23:12:30.0 & 15.82$\pm$0.05 & 16.37$\pm$0.04 & 17.11$\pm$0.06 & 14.97$\pm$0.05 & 15.45$\pm$0.03 & 16.04$\pm$0.04 & 0.12 & 1.12 & U &  &  & \\
  VVVXJ171317.68-233115.5 & 17:13:18.00 & -23:31:16.0 & 15.24$\pm$0.03 & 15.77$\pm$0.03 & 16.41$\pm$0.04 & 14.13$\pm$0.03 & 14.38$\pm$0.02 & 15.68$\pm$0.03 & 0.03 & 1.35 & U &  &  & \\
  VVVXJ171319.30-232019.4 & 17:13:19.00 & -23:20:19.0 & 16.27$\pm$0.06 & 16.96$\pm$0.05 & 17.69$\pm$0.08 & 15.23$\pm$0.06 & 16.27$\pm$0.05 & 16.89$\pm$0.06 & 0.01 & 1.40 & U &  &  & \\
  VVVXJ171323.20-240353.5 & 17:13:23.00 & -24:03:54.0 & 13.95$\pm$0.02 & 14.23$\pm$0.01 & 14.92$\pm$0.02 & 12.12$\pm$0.01 & 12.40$\pm$0.01 & 13.22$\pm$0.01 & 0.03 & 2.53 & E & 0.034 & 2MASX J17132339-2403556 & 2020yCat.1350....0G\\
  VVVXJ171323.07-233253.8 & 17:13:23.00 & -23:32:54.0 & 16.15$\pm$0.06 & 16.70$\pm$0.05 & 17.49$\pm$0.07 & 14.97$\pm$0.05 & 15.13$\pm$0.03 & 16.04$\pm$0.04 & 0.00 & 1.39 & U &  &  & \\
  VVVXJ171324.19-233914.8 & 17:13:24.00 & -23:39:15.0 & 12.77$\pm$0.01 & 13.15$\pm$0.01 & 13.84$\pm$0.01 & 11.18$\pm$0.01 & 11.36$\pm$0.00 & 11.97$\pm$0.01 & 0.03 & 2.30 & S & 0.028 & 2MASX J17132420-2339146 & 2006AJ....131.1163S\\
  VVVXJ171330.84-232224.8 & 17:13:31.00 & -23:22:25.0 & 14.54$\pm$0.02 & 14.92$\pm$0.02 & 15.65$\pm$0.03 & 13.74$\pm$0.01 & 13.90$\pm$0.01 & 14.52$\pm$0.02 & 0.43 & 1.07 & E &  &  & \\
  VVVXJ171339.27-233424.5 & 17:13:39.00 & -23:34:25.0 & 16.08$\pm$0.06 & 16.65$\pm$0.05 & 17.41$\pm$0.07 & 15.67$\pm$0.06 & 16.08$\pm$0.05 & 16.79$\pm$0.06 & 0.04 & 0.74 & U &  &  & \\
  VVVXJ171340.27-240240.4 & 17:13:40.00 & -24:02:40.0 & 15.40$\pm$0.04 & 15.91$\pm$0.03 & 16.65$\pm$0.05 & 14.35$\pm$0.03 & 14.66$\pm$0.02 & 14.96$\pm$0.03 & 0.15 & 1.39 & U &  &  & \\
  VVVXJ171345.04-232527.4 & 17:13:45.00 & -23:25:27.0 & 14.97$\pm$0.03 & 15.45$\pm$0.03 & 16.16$\pm$0.04 & 13.63$\pm$0.01 & 13.81$\pm$0.01 & 14.47$\pm$0.02 & 0.03 & 1.84 & E &  &  & \\
  VVVXJ171347.53-232243.6 & 17:13:48.00 & -23:22:44.0 & 16.14$\pm$0.06 & 16.74$\pm$0.05 & 17.46$\pm$0.07 & 15.54$\pm$0.07 & 15.83$\pm$0.04 & 16.46$\pm$0.05 & 0.01 & 0.90 & S &  &  & \\
  VVVXJ171348.04-232033.6 & 17:13:48.00 & -23:20:34.0 & 16.81$\pm$0.09 & 17.51$\pm$0.07 & 18.53$\pm$0.12 & 16.26$\pm$0.10 & 16.92$\pm$0.07 & 17.82$\pm$0.10 & 0.37 & 0.81 & U &  &  & \\
  VVVXJ171349.88-231634.1 & 17:13:50.00 & -23:16:34.0 & 15.97$\pm$0.05 & 16.63$\pm$0.05 & 17.43$\pm$0.07 & 15.12$\pm$0.05 & 15.56$\pm$0.04 & 16.31$\pm$0.05 & 0.01 & 1.13 & S &  &  & \\
  VVVXJ171349.62-224253.4 & 17:13:50.00 & -22:42:53.0 & 14.89$\pm$0.03 & 15.53$\pm$0.03 & 16.23$\pm$0.04 & 13.89$\pm$0.03 & 14.21$\pm$0.02 & 14.81$\pm$0.02 & 0.03 & 1.22 & U &  &  & \\
  VVVXJ171411.57-232646.3 & 17:14:12.00 & -23:26:46.0 & 16.06$\pm$0.06 & 16.74$\pm$0.05 & 17.53$\pm$0.07 & 15.39$\pm$0.06 & 15.93$\pm$0.04 & 16.65$\pm$0.05 & 0.15 & 0.96 & U &  &  & \\
  VVVXJ171436.50-234712.1 & 17:14:37.00 & -23:47:12.0 & 15.63$\pm$0.04 & 16.00$\pm$0.03 & 16.61$\pm$0.05 & 14.00$\pm$0.03 & 14.10$\pm$0.02 & 14.71$\pm$0.02 & 0.00 & 2.07 & S &  &  & \\
  VVVXJ171517.94-233535.7 & 17:15:18.00 & -23:35:36.0 & 13.61$\pm$0.02 & 13.97$\pm$0.01 & 14.67$\pm$0.02 & 11.93$\pm$0.01 & 12.31$\pm$0.01 & 12.87$\pm$0.01 & 0.03 & 2.53 & E &  &  & \\
  VVVXJ171541.59-225329.6 & 17:15:42.00 & -22:53:30.0 & 13.50$\pm$0.01 & 13.87$\pm$0.01 & 14.58$\pm$0.02 & 12.44$\pm$0.01 & 12.58$\pm$0.01 & 13.28$\pm$0.01 & 0.22 & 1.47 & E &  &  & \\
  VVVXJ171546.83-225731.0 & 17:15:47.00 & -22:57:31.0 & 16.09$\pm$0.06 & 16.73$\pm$0.05 & 17.48$\pm$0.07 & 14.53$\pm$0.05 & 15.16$\pm$0.03 & 15.89$\pm$0.04 & 0.14 & 2.10 & U &  &  & \\
  VVVXJ171558.55-232958.4 & 17:15:59.00 & -23:29:58.0 & 14.80$\pm$0.03 & 15.41$\pm$0.03 & 16.16$\pm$0.04 & 13.67$\pm$0.03 & 14.02$\pm$0.02 & 14.79$\pm$0.02 & 0.04 & 1.49 & U &  &  & \\
\hline
  \end{tabular}

\label{tab:trips}

\end{landscape}

\end{appendix}
\clearpage

\end{document}